\documentclass[
 nopreprint
]{jasatex}

\usepackage{color}  
\usepackage{graphicx}
\usepackage{dcolumn}
\usepackage{amsmath,amsfonts}
\usepackage{bm}

\newcommand{\beq}[1]{  \begin{equation} \label{#1} }  
\newcommand{\eeq}{     \end{equation}} 	
\newcommand{\bal}[1]{  \begin{align} \label{#1} }  
\newcommand{\rf}[1]{(\ref{#1})}
\def\bd#1{\mbox{\boldmath$\displaystyle\mathbf{#1}$} }
\def\tr{\operatorname{tr}} 
\def\dd{\operatorname{d}} 
\def\det{\operatorname{det}} 
\def\Re{\operatorname{Re}} 
\def\rev#1{\textcolor{blue}{#1}}	   
\begin{document}

\title[Diffuse wave density]{Diffuse wave density and directionality in anisotropic solids}

\author{Andrew N. Norris}
  \email{norris@rutgers.edu}
\affiliation{Mechanical and Aerospace Engineering, Rutgers University, Piscataway NJ 08854}

\date{\today}

\begin{abstract}
Several  general results are derived for diffuse waves in anisotropic solids, including   
concise expressions for the modal  density  per unit volume $d(\omega)$, and for the participation factor matrix $\bf G$.  The latter is a second order tensor which describes the orientational distribution of diffuse wave or reverberant energy, and reduces to the identity $\bf I$ under isotropy.  Calculations of $\bf G$ for a variety of example  materials   show  significant deviation  from $\bf I$ even under moderate levels of anisotropy.  

\end{abstract}

\maketitle

\section{Introduction}

 We consider how material  anisotropy effects the directional partition of reverberant or diffuse wave energy.   Diffuse waves in solids are the long time response when multiple scattering has equilibrated the energy distribution among  modes.   Preferential orientation of the root mean square particle velocity  does not arise in isotropic materials but is a characteristic of anisotropy.   Our objective is to describe this orientation effect and to quantify it in real materials.   An  ability to determine, directly or by inference, the orientational distribution of   kinetic energy density in a solid allows one to essentially ``hear'' the texture of a crystal.   We will demonstrate that the key quantity that needs to be measured  is the  autocorrelation function, or the  Green's function evaluated at its source.  By deriving an explicit formula for the autocorrelation, or the admittance matrix, we can completely describe the directional distribution of the diffuse wave energy. 

We introduce  two quantities for the description of reverberant energy in the presence of anisotropy:  the participation tensor ${\bd G}$ and the 
 modal spectral density per unit volume, $d(\omega )$.
The two are in fact  intimately related as we will see.   Under steady state time harmonic conditions the total energy of a body is equally divided between   potential and kinetic.  The latter is $\frac12 \omega^2 \int \dd V \rho |\bar{\bd u}|^2$ where $|\bar{\bd u}|$ is the \rev{root mean square} particle displacement, and assuming a uniform spatial distribution, the total energy  is $E=V \rho \omega^2|\bar{\bd u}|^2 $.   This  
 may be inverted to express the mean square displacement.  Let $\bar{u}_i = |\bar{\bd u}\cdot {\bf e}_i|$ where ${\bf e}_i$, $i=1,2,3$ is an orthonormal triad. Since $\bar{u}_1^2+\bar{u}_2^2+\bar{u}_3^2= |\bar{\bd u}|^2$ we may  write 
\beq{560}
\bar{u}^2_i = \frac{E}{3V\rho \omega^2}\,  \bar{G}_i,\qquad 
\bar{G}_i= {\bf e}_i\cdot  {\bf G}\cdot {\bf e}_i,   
\eeq
for  i=1,2,3 (no sum) where
 $ {\bf G}$ is a second order symmetric tensor satisfying
\beq{562}
\tr\, {\bf G}  = 3. 
\eeq

For isotropic materials $\bd G$ is simply the unit matrix or identity (second order) tensor. Deviations from this can occur under three  general situations: (i) If the field point is near a surface or boundary.  This was considered in detail by Weaver \cite{Weaver82}  who found expressions for the components of $\bd G$ at a free surface in terms of simple integrals, see also Egle \cite{Egle81}.   (ii) By analogy, $\bd G$ will be influenced by local inhomogeneity in the material, for instance if the field point is close to a rigid inclusion, or a void.  We will not discuss this further here.  (iii) Material anisotropy can also influence $\bd G$. Here we consider the simplest case of a field point in a homogeneous material of infinite extent.  It is expected that ${\bf G}$  displays the symmetries appropriate to the degree of anisotropy.  Thus, it is characterized by a single parameter for materials with  isotropic and cubic symmetries, and by two or three parameters for materials with lower symmetry.

The  spectral density of modes  $D$ at frequency $\omega$ in  a volume $V$  is 
 $D(\omega ) = Vd(\omega )$.   
It can be estimated as $D = \partial N/\partial \omega  \approx  V \omega^2/c^3$ by noting the total number of modes scales as $N({\bf k}) \approx  V k^3$ where $k=\omega /c$ is  typical wavenumber.  A more precise counting yields, for isotropic bodies, the well-known result 
\cite{Kittel}
\beq{86}
d(\omega ) = \frac{\omega^2}{2\pi^2} \big( \frac{2}{c_t^3} + \frac{1}{c_l^3}\big) , 
\eeq
where $c_l$ and $c_t$ are the longitudinal and transverse elastic wave speeds.   

The objective  is to derive  analogous expressions of $d(\omega)$ and ${\bd G}$ for anisotropic elastic materials.  This will be achieved by explicit calculation of the admittance tensor $\bd A$, defined in Section \ref{sec2}, combined with a general relation between  $d(\omega)$, ${\bd G}$ and ${\bd A}$.  The   spectral density and the participation tensor in the presence of material anisotropy  do not appear to have received much attention.  
Some work on the related issue of admittance in bounded anisotropic thin plate systems has appeared   \cite{Langley96}.  Weaver \cite{Weaver84}  considered isotropic plates of finite thickness and infinite lateral extent.   Tewary et al. \cite{Tewary96} derived an expression for the admittance at the free surface of an anisotropic half space as a double integral. 
Here the focus is on infinite systems, and the modal density per unit volume in this limit.  Finite structures, such as plates both thin and of finite thickness, will be considered in a separate paper. 

Our principal results are that the modal spectral density per unit volume and the participation tensor are given by 
\begin{subequations}\label{635}
\begin{align}\label{635a}
d(\omega )  =&   \frac{\omega^2}{2\pi^2}\,  \langle \tr {\bd Q}^{-3/2} \rangle  ,
\\
 {\bd G} =& 3 \frac{  \langle {\bd Q}^{-3/2} \rangle }{  \langle \tr {\bd Q}^{-3/2} \rangle }, 
 \label{635b}
 \end{align}
 \end{subequations}
 where ${\bd Q}({\bd n})$ is the acoustical or Christoffel tensor for plane waves propagating in the direction ${\bd n}$, and \rev{
 $\langle f \rangle$ is the orientation average of a function that depends on the direction, 
\beq{9051}
\langle f \rangle \equiv  \frac1{4\pi} \int\limits_{4\pi} \dd \Omega ({\bf n} )\, f( {\bd n}) . 
\eeq }
 
 In an isotropic solid \rf{635a} reduces to \rf{86} and ${\bd G} $ is simply the identity ${\bd I}$.  After deriving \rf{635}, the remainder of the paper will explore its implications, in particular the form of ${\bd G}$ is investigated, and the parameters in Table I 
 deduced.  
 \rev{
 It is interesting to note that  the material constant that determines the density of states of diffuse waves, $\tr  \langle  {\bd Q}^{-3/2} \rangle$,  also defines the Debye temperature $\Theta$ of a crystal.  Thus   (see Chapter 9 of 
 Ref.~\onlinecite{fed}), 
\beq{-00}
 \Theta = \frac{h}{k} \bigg( \frac{ 18 \pi^2 }{V_a \tr  \langle  {\bd Q}^{-3/2} \rangle } \bigg)^{1/3},
 \eeq
where $h$ is Planck's constant, $k$ is Boltzmann's constant, and $V_a$ is the volume per atom or lattice site.    Fedorov \cite{fed} provides a detailed discussion of 
$\tr  \langle  {\bd Q}^{-3/2} \rangle$ in this  context.  The emphasis in this paper is on the more general tensor   $ \langle  {\bd Q}^{-3/2} \rangle$ although connections with Fedorov's analysis will be mentioned later.  }

\begin{table}\label{table1}
\caption{The form of the participation tensor $\bd G$ for the different material symmetries.  TI, tet and trig are abbreviations for transverse isotropy, tetragonal and trigonal symmetries, respectively.  The $\bd e$ unit vectors are defined by the symmetry, while 
${\bd a}$, ${\bd b}$ and ${\bd c}$ result from averaging. The positive numbers $\alpha$, $\beta$ and $\gamma$ are constrained as indicated in order to satisfy Eq. \rf{562}.}
\begin{ruledtabular}
\begin{tabular}{cl}
${\bd G}$  & Material symmetry     \\
\hline
${\bd I}$
 & isotropic, cubic    
 \\
 $\alpha {\bd e}\otimes{\bd e}   + \beta ({\bd I} - {\bd e}\otimes{\bd e} )$
 & TI, tet, trig   $\alpha +2\beta=3$
 \\
 $\alpha {\bd e}_1\otimes{\bd e}_1 +
\beta {\bd e}_2\otimes{\bd e}_2 +
\gamma {\bd e}_3\otimes{\bd e}_3 $
 & orthotropic    $\alpha +\beta +\gamma=3 $
 \\
 $\alpha {\bd e}\otimes{\bd e} +
\beta {\bd a}\otimes{\bd a} +
\gamma {\bd b}\otimes{\bd b}$
 & monoclinic   $\alpha +\beta +\gamma=3 $
 \\
 $\alpha {\bd a}\otimes{\bd a} +
\beta {\bd b}\otimes{\bd b}+ 
\gamma {\bd c}\otimes{\bd c}$
 & triclinic   $\alpha +\beta +\gamma =3$
 \\
 \end{tabular}
\end{ruledtabular}
\end{table}

The outline of the paper is as follows.  The  admittance tensor $\bd A$ is defined and calculated  in Section \ref{sec2}, from which the main result \rf{635} follows.  Several alternative   representations of the fundamental quantity ${\bd Q}^{-3/2} $ are developed in Section \ref{sec3}.  In particular it is shown that ${\bd G}$ for transverse isotropy can be evaluated as  a single integral.   Weak anisotropy is considered in Section \ref{sec4} and numerical examples are presented in Section \ref{sec5}. 

\section{Derivation of $d$ and $\bf G$}\label{sec2}

\subsection{Admittance tensor}

The admittance $\bd A$ is  a second order tensor  defined by the average power radiated by a time harmonic point force $\bd F$ according to 
\beq{904}
\Pi = {\bf F} \cdot  {\bf A}\cdot {\bf F}. 
\eeq
Alternatively, $\bd A$ is equal to  the power expended at the source point - which is the more conventional definition of admittance, as the the inverse of drive point impedance.  The admittance is clearly related to the auto-correlation of the Green's function, and as such is a special case of the two-point cross correlation of the Green's function \cite{Wapenaar04}.  The important connection for the present purposes is  
 the relation between the radiation from a point force and the diffuse wave density \cite{Weaver85,Shorter05}.  In the present notation this becomes 
\beq{564}
{\bd A} = \frac{\pi}{12\rho} d(\omega ) {\bd G}.   
\eeq
A short derivation of \rf{564} is given in Appendix \ref{appa}.   The admittance of isotropic bodies is simply determined from Eq. \rf{86} and  ${\bd G}= {\bd I}$.  Our objective here is to calculate ${\bd A}$ for anisotropic solids, and then to use the result to determine $d(\omega )$ and $\bd G$. 

The central result for ${\bd A}$ is the following: 
The   second order symmetric admittance  tensor  of  Eq. \rf{904}  that determines the total  power radiated to infinity from the point source averaged over a period,  is  
\beq{905}
{\bd A} =   \frac{\omega^2}{8\pi \rho }\, \langle 
 {\bd Q}^{-3/2}   \rangle , 
\eeq
where  $ {\bd Q}({\bd n})$ is  the acoustical tensor, 
\beq{00}
Q_{ik}({\bf n}) = c_{ijkl}n_jn_l \quad \text{with }\quad 
c_{ijkl}= \frac1{\rho} C_{klij}.
\eeq 
The elastic moduli (stiffness)  $C_{ijkl}$  have the symmetries $C_{ijkl}= C_{klij}$ and $C_{ijkl}= C_{jikl}$, and thus have at most 21 independent elements. 
Note that ${\bd A}$ has dimensions of admittance (inverse impedance).  We next derive Eq. \rf{905} by explicitly calculating the admittance for a time harmonic point force. 

\subsection{Radiation from a point force}

The displacement resulting from a  point force ${\bf F}\cos \omega t $ at the origin is ${\bf u} ({\bf x},t) =  \Re\, \tilde{\bf u}( {\bf x}, \omega ) e^{-i\omega t}$ where $\tilde{\bf u}$ satisfies
\beq{0301}
C_{ijkl}\tilde{ u}_{k,jl} + \rho \omega^2 \tilde{ u}_i = -F_i \delta({\bf x}),
\quad -\infty \le x_1, x_2, x_3 \le \infty .
\nonumber 
\eeq
Here $\rho$ is the mass density and $\delta({\bf x})$ is the three-dimensional Dirac delta function.  
The equation of motion may be written 
\beq{0302}
{\bf Q}( \nabla ) \tilde{\bf u}  +   \omega^2 \tilde{\bf u} =  -\frac1{\rho} \delta( {\bf x} )\, {\bf F},  
\eeq
and the problem definition is completed by the requirement that the energy radiates away from the point source.

The solution to \rf{0302} in a   solid of infinite extent is well known.   For our purpose we will find  the following representation from Norris\citep[Eq. (3.22)]{norris94b}  useful for determining the admittance:
\bal{3.22}
\tilde{\bf u} 
= &\frac{1}{8\pi^2 \rho |{\bf x} |}\, \oint  \dd \theta( {\bf n} ) 
  \sum_{j = 1}^3  \frac{ {\bf q}_j\otimes {\bf q}_j }
{\lambda_j} \, {\bf F}        
 \nonumber \\ &
+
        \frac{1}{16\pi^2 \rho }\, \int\limits_{4\pi}  \dd \Omega ({\bf n} )
  \sum_{j = 1}^3
\frac{i k_j}{ \lambda_j}\,    {\bf q}_j\otimes {\bf q}_j\, {\bf F} 
e^{ik_j  {\bf n} .{\bf x}  }  .    
\end{align}
Here  $\lambda_1, \lambda_2, \lambda_3$ are the eigenvalues and ${\bd q}_1,{\bd q}_2,{\bd q}_3$ the eigenvectors of ${\bd Q}( {\bd n})$, 
which then has the  spectral decomposition 
\beq{1}
{\bd Q}( {\bd n}) =  \lambda_1 {\bd q}_1  \otimes {\bd q}_1
+ \lambda_2 {\bd q}_2  \otimes {\bd q}_2
+ \lambda_3 {\bd q}_3  \otimes {\bd q}_3.     
\eeq
Also, $k_j =  {\omega}/{\lambda_j^{1/2}}$ are the wavenumbers of the three distinct branches of the slowness surface defined by the eigenvectors.  
The first integral in \rf{3.22} is around the  unit circle  formed by the
intersection of the plane ${\bf n }\cdot {\bf x} = 0$ with the unit ${\bf n}
-$sphere. This is just  the static Green's function of elasticity \cite{norris94b}. 
The important dynamic quantity is the second integral which is evaluated over the 
 sphere $ \{| {\bf n}|=1\}$.  In order to make this more apparent, we rewrite \rf{3.22} as 
\beq{3.224}
\tilde{\bf u} 
= \tilde{\bf u}|_{\omega = 0}  
+
        \frac{i\omega}{4\pi \rho }\,  \sum_{j = 1}^3 \big\langle    e^{ik_j  {\bf n} .{\bf x}  } 
\frac{{\bf q}_j \otimes {\bf q}_j }{\lambda_j^{3/2}   } \big\rangle \, {\bf F} 
 ,    
\eeq
and note for future reference that the first term on the right hand side is real valued. 

The average power radiated per  period is equal to the power expended by the force
\beq{97}
\Pi = \lim_{{\bf x} \rightarrow 0} \frac{\omega}{2\pi}\,  \int\limits_0^{ 2\pi /\omega}  \dd t\, \cos \omega t \, {\bf F} \cdot {\bf v}({\bf 0},t), 
\eeq
where ${\bf v}({\bf x},t)=  \Re\,\big( -i\omega \tilde{\bf u}( {\bf x}, \omega ) e^{-i\omega t}\big) $ is the particle velocity.  Thus, 
\beq{971}
\Pi =   \frac{\omega^2}{8\pi \rho }\,  
\, \sum_{j = 1}^3 \langle 
\frac{1}{ \lambda_j^{3/2}}\,    ( {\bf q}_j \cdot {\bf F} )^2  \rangle . 
\eeq
The spectral decomposition  \rf{1} implies that 
\beq{2}
  \lambda_1^{-3/2} {\bd q}_1  \otimes {\bd q}_1
+ \lambda_2^{-3/2} {\bd q}_2  \otimes {\bd q}_2
+ \lambda_3^{-3/2} {\bd q}_3  \otimes {\bd q}_3  = {\bd Q}^{-3/2},
\nonumber 
\eeq
which together with Eq. \rf{904}  proves the main result \rf{905}. 

The scalar $d(\omega)$ and the tensor $\bf G$ are defined such that their product is $12 \rho /\pi$ times the admittance $\bd A$, see \rev{Eqs. \rf{86}, \rf{635}, \rf{564} and \rf{905}}.  This defines $d$ and $\bf G$ to within a constant, which is determined uniquely by the constraint $\tr {\bd G}=3$. We therefore  obtain the general results of Eq. \rf{635}.  
As discussed, $d$ is the generalization of the classical density of states per unit volume, \rf{86} for isotropic solids, and the participation factor tensor $\bd G$ describes the directional distribution of the energy at a point. While it is convenient to consider them separately, $d$ and $\bd G$ are both defined by the averaged tensor
$ \langle {\bd Q}^{-3/2} \rangle$, which will be the focus of the remainder of the paper.  

Before considering the properties of $d$ and ${\bd G}$ we  note that the isotropic modal density of states  follows immediately from \rf{635a}.  Starting with the acoustical tensor for an isotropic solid, 
\beq{090}
{\bd Q}  ( {\bd n})  = c_l^2 {\bd n}\otimes {\bd n} +  c_t^2 ({\bd I} - {\bd n}\otimes {\bd n}), \quad\text{isotropy},
\eeq
we have   ${\bd Q}^{-3/2} =   c_l^{-3} {\bd n}\otimes {\bd n} +  c_t^{-3} ({\bd I} - {\bd n}\otimes {\bd n})$.  Then using the fact that $
\langle {\bd n}\otimes {\bd n} \rangle  = \frac{1}{3} {\bd I}$ 
it follows that 
\beq{9060}
\langle {\bd Q}^{-3/2}\rangle = \frac13(c_l^{-3} + 2c_t^{-3}  ) {\bd I}.
\eeq
Hence, the density of states per unit volume is $d=  \frac{\omega^2}{2\pi^2}(c_l^{-3} + 2c_t^{-3}  ) ^{-1}$, in agreement with the well known identity \rf{86}, and ${\bd G} = {\bd I}$, as expected. 

\section{${\bf Q}^{-3/2}$ and related quantities}\label{sec3}

The key quantity is the tensor ${\bd Q}^{-3/2}$ and its directional average.  In practice, this may be evaluated numerically without difficulty.  It is however useful to examine semi-explicit forms for the tensor, both for general anisotropy and for specific symmetries, particularly the case of  transverse isotropy.  We begin with two alternative 
and general formulations based on the spectral properties and the invariants of the acoustical tensor. 

\subsection{General representations for arbitrary anisotropy}

\subsubsection{A method based on invariants}

Functions of a positive definite tensor can be simplified using the Cayley-Hamilton formula for the tensor, which for ${\bd Q}$ is 
\beq{-33}
{\bd Q}^3 - I_1 {\bd Q}^2 +I_2 {\bd Q} - I_3 {\bd I}= 0. 
\eeq
The  principal positive invariants of ${\bd Q}$ are  
\beq{4}
 I_1 = \tr {\bd Q}, \quad
   I_2 =  \frac12  (\tr {\bd Q})^2 - \frac12  \tr {\bd Q}^2 ,
   \quad
  I_3  = \det {\bd Q} . 
\eeq
Based on these fundamental properties, it can be shown that
\bal{-21}
 {\bd Q}^{-3/2}  =&   \big[
 (  I_1 I_3  +  i_1 i_3I_2 +i_2I_3 ) ({\bd Q}^2 -I_1{\bd Q} + I_2 {\bd I})
 \nonumber \\ &
 + i_1 i_3 I_3 ( {\bd Q} - I_1 {\bd I})  - I_3^2 {\bd I}\big] /\big[   ( i_1 i_2- i_3)I_3^2 \big],  
 \end{align}
  where  $i_1 $, $i_2 $ and $i_3 $ are the positive invariants of ${\bd Q}^{1/2}$ which can be expressed as functions of the invariants $I_1$,$I_2$ and $I_3$, see below.  Details of the derivation of 
   \rf{-21} are given  in Appendix B.  
   
  The appealing feature of Eq. \rf{-21} for $ {\bd Q}^{-3/2} ( {\bd n})$  is that it only involves powers of $\bd Q$, its three invariants, and the additional invariants $i_1$, $i_2$ and $i_3$.   These are  
   related to  $I_1 $, $I_2 $ and $I_3 $ by \cite{Hoger84,Norris07a}
  \beq{62}
   i_1^2-2i_2= I_1,
\qquad
 i_2^2 - 2i_1 i_3= I_2, 
\qquad
  i_3^2= I_3. 
\eeq
The last implies $i_3 = I_3^{1/2}$, while expressions for  
$i_1$ and $i_2$  are given by Hoger and Carlson \cite{Hoger84} and by Norris \cite{Norris07a}.  For instance  \citep{Norris07a},
 \begin{subequations}
\begin{align}\label{13}
i_1 = &\sqrt{I_1-\beta +2 \sqrt{I_3/\beta} } +\sqrt{\beta} ,
\\
i_2=  &\sqrt{I_2 - I_3/\beta+2 \sqrt{I_3 \beta} } +\sqrt{I_3/\beta} ,
\\
i_3= &\sqrt{I_3},
\end{align}
\end{subequations}
where $\beta$ is any eigenvalue of $\bd Q$, e.g.      
\begin{subequations}
\begin{align}\label{15}
 \beta = & \frac13 \big( I_1 + \big[(\xi+\sqrt{\xi^2-( I_1^2 -3I_2)^3} \big]^{1/3}
 \nonumber \\ & 
 													\,\,			+ \big[(\xi-\sqrt{\xi^2-( I_1^2 -3I_2)^3} \big]^{1/3} \big),
 \\
 \xi = &   \frac12 (2 I_1^3- 9 I_1I_2+27 I_3).
\end{align}
\end{subequations}
Note that \cite{Carroll04} $i_1 i_2 - i_3 = \det( i_1 {\bd I} - {\bd Q}^{1/2}) > 0$.

Taking  the trace of Eq. \rf{-21} gives 
\beq{-4}
\tr {\bd Q}^{-3/2}
= \frac{      (I_1+ i_2)I_2 I_3  + (I_2^2 -2 I_1 I_3 ) i_1 i_3 - 3 I_3^2
 } {  ( i_1 i_2 - i_3) I_3^2 }.
\eeq
This  quantity, when averaged over all orientations, gives the density of states function $d(\omega)$ of Eq. \rf{635a}.   Hence $d$ can be calculated from the invariants $\bd Q$ and the derived invariants $i_1$, $ i_2$, $ i_3$. 

\subsubsection{A spectral representation}

The second form for ${\bd Q}^{-3/2}$ is based on the spectral decomposition \rf{2}. 
The latter  can be expressed in a form that does not explicitly involve the eigenvectors, 
\beq{29}
{\bd Q}^{-3/2}
=  \lambda_1^{-3/2} {\bd N}(\lambda_1) 
+ \lambda_2^{-3/2} {\bd N}(\lambda_2)
+ \lambda_3^{-3/2} {\bd N}(\lambda_3) . 
\eeq
The second order tensors  ${\bd N}(\lambda_j)$, which  are alternative expressions for the dyadics formed by the eigenvectors, 
${\bd q}_j\otimes {\bd q}_j$, can be  expressed in terms of $\bf Q$ using Sylvester's formula
\beq{291}
{\bd N}(\lambda , {\bd n})  = \frac{
\lambda {\bd Q}^2 + (\lambda - I_1) \lambda  {\bd Q} + I_3{\bd I} }
{\lambda^3 + (\lambda - I_1) \lambda^2 + I_3}. 
\eeq
  The identity \rf{29} is derived in  Appendix \ref{appb}. 

 Calculation of \rf{29} requires knowledge of the 
three eigenvalues, which are zeros of the characteristic polynomial defined by Eq. \rf{-33}, 
\beq{+33}
p(\lambda ) = \lambda ^3 - I_1 \lambda ^2 +I_2 \lambda  - I_3 .
\eeq
The eigenvalues $\{ \lambda_1,\lambda_2,\lambda_3 \}$ can be expressed in terms of the invariants as 
\beq{292}
\{ \beta, \frac12(I_1 - \beta) \pm  \frac12 \sqrt{
(I_1 - \beta)^2 - 4 I_3/\beta }\} , 
\eeq
where  $\beta$ is defined in \rf{15}. Every \cite{Every80} derived alternate closed-form expressions based on the trigonometric solution of the characteristic cubic.   The alternative version of Eq. \rf{-4} is 
\beq{-41}
\tr {\bd Q}^{-3/2}
= \lambda_1^{-3/2}+ \lambda_2^{-3/2}+ \lambda_3^{-3/2},
\eeq
which is the starting point for Fedorov's calculation \cite{fed} of the trace. 

\subsection{Transverse isotropy}

Transverse isotropy or hexagonal symmetry is an important class of anisotropy.  It occurs in many practical circumstances, whether from layering in the earth to laminated composite materials, or from underlying crystal structure. It is  the highest symmetry for which the participation factor tensor is not the identity, \rev{since ${\bf G}={\bf I}$ under isotropy and cubic material symmetry}.  We now demonstrate that the evaluation of $d$ and $\bf G$ may be reduced to the evaluation of two single integrals, one for 
$\langle \tr {\bd Q}^{-3/2} \rangle $ and one for the parameter $\alpha$ that defines $\bf G$, see Table I. 

Transversely isotropic solids have five independent moduli: 
$c_{11}=c_{22}$, $c_{33}$, $c_{12}$, $c_{13}=c_{23}$, $c_{44}=c_{55}$, $c_{66}=\frac12(c_{11}-c_{12})$.
Let ${\bd e}$ be the  axis of symmetry.  
The SH slowness decouples to give
\beq{701}
{\bd Q} = \lambda_3 ( {\bd n} \cdot{\bd e}) \, {\bd q}_3\otimes {\bd q}_3 + {\bd Q}_\perp, 
\eeq
where \cite[p. 95]{Musgrave} 
\beq{7023}
\lambda_3 ( {\bd n} \cdot{\bd e}) = c_{66} + (c_{44} - c_{66}) ({\bd n} \cdot{\bd e})^2, 
\eeq
and ${\bd q}_3 = {\bd e}  \wedge {\bd n}/|{\bd e}  \wedge  {\bd n} |$. The 2-dimensional symmetric tensor ${\bd Q}_\perp$
is \cite{Musgrave}
\begin{align}\label{7014} 
{\bd Q}_\perp = & [ c_{44} + (c_{33} - c_{44})({\bd n} \cdot{\bd e})^2]  {\bd e}\otimes {\bd e} 
\nonumber \\ &
+[ c_{11} + (c_{44} - c_{11})({\bd n} \cdot{\bd e})^2]  {\bd d}\otimes {\bd d} 
\nonumber \\ & 
+ (c_{13} + c_{44}){\bd n} \cdot{\bd e} \sqrt{1- ({\bd n} \cdot{\bd e})^2} [ {\bd d}\otimes {\bd e} + {\bd e}\otimes {\bd d}],\nonumber
\end{align}
where ${\bd d} = {\bd e}\wedge {\bd q}_3$. 
Replacing ${\bd n} \cdot{\bd e}$ by the integration parameter $\xi$, it follows that 
\beq{7011}
\langle \lambda_3^{-3/2} {\bd q}_3\otimes {\bd q}_3 \rangle
=   \frac12 \int\limits_0^1\dd \xi \, \lambda_3^{-3/2} ( \xi)\, {\bd I}_\perp ,  
\eeq
where  ${\bd I}_\perp$ projects onto the plane perpendicular to ${\bd e}$, 
\beq{-999}
{\bd I}_\perp = {\bd I}- {\bd e}\otimes {\bd e}. 
\eeq
It remains to consider the orientational average of ${\bd Q}_\perp^{-3/2}$. 

The   tensor ${\bd Q}_\perp$ satisfies a quadratic Cayley-Hamilton equation
\beq{702}
{\bd Q}_\perp^2 - J_1  {\bd Q}_\perp + J_2 {\bd I}_\perp = 0,
\eeq
with $J_1 = \tr {\bd Q}_\perp = \lambda_1+\lambda_2$ and  $J_2 = \det {\bd Q}_\perp = \lambda_1\lambda_2$.  Similarly, the Cayley-Hamilton equation for the square root is 
\beq{703}
({\bd Q}_\perp^{1/2})^2 - j_1  {\bd Q}_\perp^{1/2} + j_2 {\bd I}_\perp = 0,
\eeq
where $j_1 = \tr {\bd Q}_\perp^{1/2} $ and   $j_2  = \det {\bd Q}_\perp^{1/2}$ satisfy 
$J_1= j_1^2-2 j_2$, $J_2= j_2^2$, and are therefore related to $J_1$ and $J_2$
by
$j_1  = \sqrt{J_1 + 2 \sqrt{J_2}}$, $j_2  = \sqrt{J_2}$. 
Using Eqs. \rf{702} and \rf{703}, respectively, leads to the identities
\begin{subequations}\label{705}
\begin{align}
{\bd Q}_\perp^{-2} =  &J_2^{-2} \big[ (J_1^2 - J_2) {\bd I}_\perp  - J_1{\bd Q}_\perp  \big],
\\
{\bd Q}_\perp^{1/2} =  &j_1^{-1} ( {\bd Q}_\perp +j_2 {\bd I}_\perp) . 
\end{align}
\end{subequations}
Multiplication of these and further use of  \rf{702} leads to 
\beq{706}
{\bd Q}_\perp^{-3/2} = \frac{1}{j_1j_2J_2}\big[ (J_1+j_2)(J_1 {\bd I}_\perp - {\bd Q}_\perp) - J_2 {\bd I}_\perp\big]. 
\eeq

Again using $\xi = {\bd n} \cdot{\bd e}$, we have 
\bal{707}
&\langle \tr {\bd Q}^{-3/2}   \rangle
= 
\nonumber \\ 
&\quad \int\limits_0^1\dd \xi \, \big[  J_2^{-3/2} (J_1-\sqrt{J_2})\sqrt{J_1 + 2 \sqrt{J_2}}  + 
\lambda_3^{-3/2} (\xi)
\big] , \nonumber
\end{align}
 and  from Table I, 
 \beq{77}
 \alpha = \frac{3}{ 
\langle \tr {\bd Q}^{-3/2}   \rangle
}
\int\limits_0^1 \dd \xi 
\frac{( J_1 +  \sqrt{J_2})(J_1 - {\bd e} \cdot{\bd Q}_\perp\cdot {\bd e}) - J_2}
{ J_2^{3/2} \sqrt{ J_1 + 2 \sqrt{J_2}} }  .\nonumber
\eeq
The modal density parameter $\langle \tr {\bd Q}^{-3/2}   \rangle$ and the scalar 
$\alpha$ that defines the participation tensor can therefore be expressed as single integrals, which follow  from the above results and  Eqs. \rf{701} through \rf{7011}, as
\begin{widetext}
\begin{subequations}\label{708}
\begin{align}\label{708a}
\langle \tr {\bd Q}^{-3/2}   \rangle
=& \int\limits_0^1\dd \xi \, \bigg[  
\frac{( a+b\xi^2 -\sqrt{d+e\xi^2+f\xi^4 })\sqrt{a+b\xi^2  + 2 \sqrt{d+e\xi^2+f\xi^4}} }  
{ ( d+e\xi^2+f\xi^4)^{3/2} } 
+ \frac{1}{[ c_{66} + (c_{44} - c_{66}) \xi^2]^{3/2} } 
\bigg] , 
\\
\alpha = &\frac{3}{ 
\langle \tr {\bd Q}^{-3/2}   \rangle
}
\int\limits_0^1  \dd \xi  
\bigg[   \frac{ (c_{11}+c \xi^2) ( a+b\xi^2  +  \sqrt{d+e\xi^2+f\xi^4}) - (d+e\xi^2+f\xi^4) }{ 
(d+e\xi^2+f\xi^4)^{3/2}    \sqrt{  a+b\xi^2  + 2 \sqrt{d+e\xi^2+f\xi^4} } }
  \bigg] ,\label{708b}
\end{align}
\end{subequations}
\end{widetext}
where
\begin{align}\label{709}
&a=c_{11}+c_{44},\qquad 
b=c_{33}-c_{11}, 
\nonumber \\
&c= c_{44} -  c_{11} , \qquad 
d =  c_{11} c_{44},
\nonumber \\
&
e = c_{11} c_{33} - c_{13}^2  - 2c_{44} (c_{11}+ c_{13}) , 
\nonumber \\
&
f =  - c_{11} c_{33} + c_{13}^2 + c_{44} (c_{11}+c_{33}+ 2c_{13}).\nonumber
\end{align}

\section{Weak anisotropy}\label{sec4}

Although  the general expressions for the modal density $d$ and the participation tensor $\bf G$  are not difficult to compute, it is often the case that the medium is to a first approximation isotropic, and appropriate approximations can be made.  The state of small or \emph{weak anisotropy}  is defined relative to a background isotropic medium, and it is important to select the latter properly.  
In this Section we calculate $d$ and $\bf G$ in the presence of weak anisotropy.  
\rev{ Fedorov \cite{fed} provides a detailed analysis of the expansion of 
 $\tr \langle {\bd Q}^{-3/2} \rangle$ to arbitrary orders in the perturbation parameter.  Our emphasis is more on obtaining estimates of the tensor $\langle {\bd Q}^{-3/2} \rangle$, which is not discussed explicitly by Fedorov.  }
 We begin with a description of the comparison isotropic moduli and then proceed to calculate the first two terms in a perturbation series for $d$ and $\bf G$. 

\subsection{Background isotropic moduli}

Regardless of the level of the anisotropy it is always possible to define a unique set of isotropic moduli which minimize the Euclidean distance between the exact set of moduli and the equivalent isotropic moduli \cite{Norris05g}. This procedure is equivalent  to requiring that the mean square Euclidean difference in the slowness surfaces is minimal \cite{fed,Norris05g}.  Thus, let the  background  isotropic moduli be 
\beq{000}
 c^{(0)}_{ijkl} = c_l^2 \delta_{ij}\delta_{kl}  +  c_t^2 
 (\delta_{ik}\delta_{jl}+ \delta_{il}\delta_{jk}- 2\delta_{ij}\delta_{kl}),
\eeq
where $c_l$ and $c_t$ are the effective longitudinal and transverse wave speeds.  These are defined  by simultaneously minimizing the quantity  $\langle | {\bd Q} - {\bd Q}_0 |^2 \rangle $ with respect to both 
$c_l$ and $c_t$, where ${\bd Q}_0 ({\bf n})$ is defined by the moduli $c^{(0)}_{ijkl}$. 
The unique solution is 
\beq{7901}
c_l^2 = \frac13\, \tr\, {\bf C}_l, \qquad c_t^2 = \frac13\, \tr\, {\bf C}_t, 
\eeq
where the second order tensors of reduced moduli are 
 \beq{790}
 C_{l,ij} = \frac{2}{5}c_{ikjk}+ \frac{1}{5}c_{ijkk},\qquad 
 C_{t,ij} = \frac3{10} c_{ikjk} - \frac1{10}\, c_{ijkk} .
\eeq
The background Lam\'e moduli $\lambda$ and $\mu$ are obtained  using $c_l^2 =(\lambda + 2\mu) /\rho$ and $c_t^2 =  \mu /\rho$.  The elements of ${\bf C}_l$ and ${\bf C}_t$ follow from 
\begin{align}
 c_{ijkk} = &\begin{pmatrix}
 c_{11} +  c_{12} +  c_{13} &  c_{16} +  c_{26} +  c_{36} &  c_{15} +  c_{25} +  c_{35} \\
 c_{16} +  c_{26} +  c_{36} &  c_{12} +  c_{22} +  c_{23} &  c_{14} +  c_{24} +  c_{34} \\
 c_{15} +  c_{25} +  c_{35} &  c_{14} +  c_{24} +  c_{34} &  c_{13} +  c_{23} +  c_{33} 
\end{pmatrix},
\nonumber \\
c_{ikjk} = &\begin{pmatrix}
 c_{11} +  c_{55} +  c_{66} &  c_{16} +  c_{26} +  c_{45} &  c_{15} +  c_{46} +  c_{35} \\
 c_{16} +  c_{26} +  c_{45} &  c_{22} +  c_{44} +  c_{66} &  c_{24} +  c_{34} +  c_{56} \\
 c_{15} +  c_{46} +  c_{35} &  c_{24} +  c_{34} +  c_{56} &  c_{33} +  c_{44} +  c_{55} 
\end{pmatrix}.\nonumber
\end{align}

\subsection{Perturbation analysis}
Let 
\beq{3410}
c_{ijkl} = c_{ijkl}^{(0)} +  \varepsilon c_{ijkl}^{(1)}, 
\eeq
where the nondimensional parameter $\varepsilon$  is introduced only to simplify the perturbation analysis.  In practice $\varepsilon$ is  set to unity on the assumption that the additional moduli $c_{ijkl} - c_{ijkl}^{(0)}$ are small in comparison with the isotropic background.  

We  seek  expansions in  powers of the small parameter $\varepsilon$.  The key quantity 
${\bd Q}^{-3/2} $ will be determined as the product of ${\bd Q}^{-2}$ and ${\bd Q}^{1/2}$.  Based on \rf{3410}, the 
acoustical tensor is 
\beq{341}
{\bd Q} = {\bd Q}_0 + \varepsilon {\bd Q}_1,   
\eeq
 and simple perturbation gives
\beq{342}
  {\bd Q}^{-2}  = {\bd Q}_0^{-2} - \varepsilon \big({\bd Q}_0^{-2}{\bd Q}_1{\bd Q}_0^{-1} 
  + {\bd Q}_0^{-1}{\bd Q}_1{\bd Q}_0^{-2} \big)
+\text{O}(\varepsilon^2).
\nonumber
\eeq
  Let
\beq{343}
{\bd Q}^{1/2} = {\bd Q}_0^{1/2} + \varepsilon {\bd S}_1+\text{O}(\varepsilon^2) , \nonumber
\eeq
then ${\bd S}_1$ satisfies 
\beq{344}
 {\bd Q}_0^{1/2}{\bd S}_1 + {\bd S}_1 {\bd Q}_0^{1/2}  = {\bd Q}_1 .  
\eeq

In order to calculate ${\bd Q}^{-2}$ and also the  square root of ${\bd Q}$, we now use the fact that the leading order 
moduli $c_{ijkl}^{(0)}$ are isotropic.
The explicit form of ${\bd Q}_0^{1/2}$  follows from Eq.  \rf{090} and the identity
\beq{345}
{\bd Q}_0^m    = c_l^{2m}  {\bd n}\otimes {\bd n} +  c_t^{2m} {\bd P},
\eeq
where $m$ is any real number and 
${\bd P}  = {\bd I} - {\bd n}\otimes {\bd n}$. \rev{
 Equation \rf{344} can be solved by observing that 
 ${\bd Q}_1$ may  be partitioned 
  ${\bd Q}_1 = {\bd Q}_1^{(1)} + {\bd Q}_1^{(2)} + {\bd Q}_1^{(3)} $ where  
   ${\bd Q}_1^{(1)} = {\bd n}\cdot {\bd Q}_1 \cdot {\bd n}) {\bd n}\otimes {\bd n}$, 
${\bd Q}_1^{(2)} = {\bd P} {\bd Q}_1  {\bd P}$ and 
 ${\bd Q}_1^{(3)} = 
 {\bd P} {\bd Q}_1\cdot {\bd n} \otimes {\bd n} + {\bd n}\otimes {\bd P} {\bd Q}_1\cdot {\bd n}$.  
 Assuming a solution of the form 
 ${\bd S}_1 = p_1  {\bd Q}_1^{(1)}+p_2  {\bd Q}_1^{(3)}+p_3  {\bd Q}_1^{(3)}$, the coefficients can be determined easily from    Eq. \rf{344}, i.e. 
\beq{346}
 {\bd S}_1   = \frac{1}{2c_l}{\bd Q}_1^{(1)} 
 + \frac{1}{2c_t} {\bd Q}_1^{(2)}
 +\frac{1}{c_l+c_t } {\bd Q}_1^{(3)} .
\eeq }
Combining the asymptotic expansions for ${\bd Q}^{-2}$ and ${\bd Q}^{1/2}$ gives  
\beq{347}
{\bd Q}^{-3/2} = {\bd Q}_0^{-3/2} + \varepsilon {\bd V}_1 +\text{O}(\varepsilon^2) , 
\eeq
where
\begin{align}
{\bd V}_1 = &
{\bd Q}_0^{-2}  {\bd S}_1 - {\bd Q}_0^{-2}{\bd Q}_1{\bd Q}_0^{-1/2}  - {\bd Q}_0^{-1}{\bd Q}_1{\bd Q}_0^{-3/2} 
 \nonumber \\
 = & - \frac{3}{2c_t^5}   {\bd Q}_1     - 
 \big[ \frac{(c_l^2+c_t^2+c_lc_t)}{c_l^3c_t^3(c_l+c_t) }  - \frac{3}{2c_t^5}\big]
 \nonumber \\ &  \qquad \times 
 \big[  {\bd Q}_1\cdot {\bd n} \otimes {\bd n} + {\bd n}\otimes  {\bd Q}_1\cdot {\bd n} \big]
  \nonumber \\
   & + 
 \big[ 2\frac{(c_l^2+c_t^2+c_lc_t)}{c_l^3c_t^3(c_l+c_t) } 
 -  \frac{3}{2c_l^5} -  \frac{3}{2c_t^5} \big] ( {\bd n}\cdot {\bd Q}_1 \cdot {\bd n}) {\bd n}\otimes {\bd n} .
 \nonumber
\end{align}
The orientational average $\langle {\bd Q}^{-3/2}\rangle$ can then be effected using the identities  
\begin{align}
\langle n_in_jn_kn_l \rangle  = &\frac{1}{15} (\delta_{ij} \delta_{kl}+\delta_{ik} \delta_{jl}+\delta_{il} \delta_{jk})
\nonumber \\ 
\equiv & K_{ijkl},
\nonumber \\
\langle n_in_jn_kn_l n_pn_q \rangle  = &\frac1{7}( \delta_{ij}K_{klpq} +
\delta_{ik}K_{jlpq} +
\nonumber \\ &
\delta_{il}K_{kjpq} +
\delta_{ip}K_{kljq} +
\delta_{iq}K_{klpj} ). 
\nonumber
\end{align} \rev{
The resulting expressions for $\langle {\bd Q}^{-3/2}\rangle$ is
\begin{align}\label{350}
&\langle {\bd Q}^{-3/2} \rangle _{ij} =  \frac13 \big( \frac{2}{c_t^3} + \frac{1}{c_l^3}\big) \delta_{ij} + \varepsilon  \bigg\{
- \frac{1}{2c_t^5} c^{(1)}_{ikjk} 
\nonumber \\
& \qquad 
 - \frac{2}{15} \big[ \frac{(c_l^2+c_t^2+c_lc_t)}{c_l^3c_t^3(c_l+c_t) }  - \frac{3}{2c_t^5}\big] 
 ( c^{(1)}_{ijkk} + 2 c^{(1)}_{ikjk}  )
 \nonumber \\
   & \qquad  +  \frac1{105}\big[ 2\frac{(c_l^2+c_t^2+c_lc_t)}{c_l^3c_t^3(c_l+c_t) } 
 -  \frac{3}{2c_l^5} -  \frac{3}{2c_t^5} \big]
 \nonumber \\
   & \qquad \times  
 \big[  \delta_{ij} ( c^{(1)}_{kkll} + 2 c^{(1)}_{klkl}  ) +4( c^{(1)}_{ijkk} + 2 c^{(1)}_{ikjk}  )\big] 
\bigg\}+\text{O}(\varepsilon^2) . 
\nonumber 
 \end{align} }

\begin{figure}
				\begin{center}	
				\includegraphics[width=3.6in , height=2.6in 					]{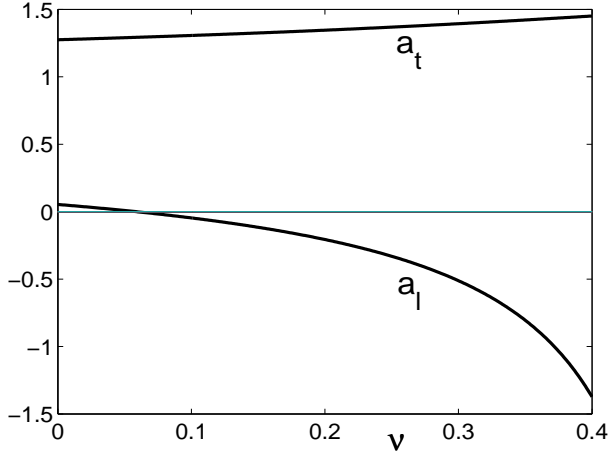} 
	\caption{The non-dimensional parameters $a_l$ and $a_t$ as a function of the Poisson's ratio $\nu$.}
		\label{fig1} \end{center}  
	\end{figure}

We note that both $  c^{(1)}_{iijj}$ and $  c^{(1)}_{ijij}$ vanish by virtue  of the choice of the background isotropic moduli.  This implies that the trace of $\langle {\bd Q}^{-3/2} \rangle$ differs from the isotropic approximant only at the second order of anisotropic perturbation,
\beq{792}
\tr \langle {\bd Q}^{-3/2} \rangle = \frac{2}{c_t^3} + \frac{1}{c_l^3}
+\text{O}(\varepsilon^2).
\eeq \rev{ 
This is  in agreement with Fedorov \cite{fed} who also provides explicit forms for the higher order terms; for instance, the expansion for cubic crystals up to fourth order in the perturbation is given by Eqs. (50.12) - (50.14) of 
Ref.~\onlinecite{fed}. 
The leading order approximation of Eq. \rf{792} when combined with  the identity \rf{635b},   gives
\begin{align}
G_{ij} = &  \delta_{ij} -  \varepsilon   \big( \frac{2}{c_t^3} + \frac{1}{c_l^3}\big)^{-1} \bigg\{
 \frac{3}{2c_t^5} c^{(1)}_{ikjk} 
 \nonumber \\ &
 + \frac{3}{35} \big[ 2\frac{(c_l^2+c_t^2+c_lc_t)}{c_l^3c_t^3(c_l+c_t) } + \frac{2}{c_l^5} - \frac{5}{c_t^5}\big] 
  \nonumber \\ & \times
 ( c^{(1)}_{ijkk} + 2 c^{(1)}_{ikjk}  )
\bigg\}+\text{O}(\varepsilon^2) .  \nonumber
 \end{align}
Ignoring terms of order $\varepsilon^2$ and then setting $\varepsilon \rightarrow 1$ yields the leading order  approximation to  the participation tensor as 
 \beq{357}
{\bd G} \approx   {\bd I} + a_l (    {\bd I} -  c_l^{-2}{\bd C}_l) 
 +  a_t(     {\bd I} -  c_t^{-2}{\bd C}_t )   , 
 \eeq }
 where the non-dimensional coefficients are 
 \begin{subequations}
 \bal{358}
a_l=&  \frac{6}{7(2+\kappa^{-3} )}
  \big(  \frac{1}{\kappa^3} +\frac{1}{\kappa} -\frac{1}{ \kappa +1 } +1 - \frac{3}{4}\kappa^2
  \big) ,
 \\
a_t =& \frac3{2+\kappa^{-3}}, 
\end{align}
\end{subequations}
and 
\beq{44}
\kappa \equiv \frac{c_l}{c_t}. 
  \eeq
Figure \ref{fig1} shows $a_l$ and $a_t$ as functions of the Poisson's ratio $\nu$, using 
$\kappa^2 = {2(1-\nu)}/(1-2\nu)$.  Note that $1.27 \ldots < a_t < 3/2$ for $0 < \nu < 1/2$
while \rev{$a_l\approx -\frac{9}{28}(1-2\nu)^{-1}$ } as $\nu \rightarrow 1/2$.  

 \subsection{Transversely isotropic materials}
As an example of the general perturbation approach, we consider the particular case of TI materials. 
We take  the axis of symmetry ($\bd e$ in Section \ref{sec3}) in the $3-$direction, so that  
\begin{align}
 c_{ijkk} = &\begin{pmatrix}
 c_{11} +  c_{12} +  c_{13} &  0 &  0 \\
 0 &  c_{11} +  c_{12} +  c_{13} &  0 \\
 0 &  0 &  c_{33} +  2c_{13}  
\end{pmatrix},
\nonumber \\
c_{ikjk} = &\begin{pmatrix}
 c_{11} +  c_{44} +  c_{66} &  0 &  0 \\
 0 &  c_{11} +  c_{44} +  c_{66} &  0 \\
 0 &  0 &  c_{33} +  2c_{44}  
\end{pmatrix}, 
\nonumber
\end{align}
where  $c_{66} = \frac12 (c_{11}- c_{12})$. The  wave speeds in the background isotropic medium are then, 
\begin{subequations}\label{4922}
\begin{align}
 c_l^2 &= \frac1{15} ( 8 c_{11} + 3 c_{33} + 4c_{13} + 8 c_{44}),
\\
c_t^2 &=   \frac1{30} ( 2 c_{11} + 2 c_{33} - 4c_{13} + 12 c_{44}+10c_{66}). 
\end{align}
\end{subequations}
According to Table I 
the participation tensor is defined by a single parameter, $\alpha$, which to leading order is unity.  Let   
\beq{-943}
\alpha = 1 - 2\beta , 
\eeq
so that 
 \beq{-99}
{\bd G} =  \begin{pmatrix}
1+ \beta &  0 &  0 \\
 0 &  1+ \beta &  0 \\
 0 &  0 &  1-2 \beta 
\end{pmatrix}. 
\eeq
Applying the general perturbation theory we find that the leading order  correction to the isotropic participation tensor is given by 
\bal{-98}
\beta  =& \frac{a_l}{15 c_l^2} ( -4c_{11} + 3 c_{33} + c_{13} + 2 c_{44}) 
\nonumber \\ & 
  + \frac{a_t}{30 c_t^2} ( -c_{11} + 2 c_{33} - c_{13} + 3 c_{44}- 5 c_{66}) 
 , 
\end{align}
where $a_l$ and $a_t$ are defined in \rf{358}.    

Thomsen's anisotropy parameters \cite{Thomsen86} $\epsilon, \gamma, \delta$  provide a means to characterize weakly anisotropic TI materials.  The parameters are defined $\epsilon = (c_{11}- c_{33})/(2c_{33})$, $\delta  = [ (c_{13} + c_{44})^2 - (c_{33}- c_{44})^2]/[ 2c_{33}(c_{33} - c_{44})]$, $\gamma = (c_{66}- c_{44})/(2c_{44})$,  and are commonly used in geophysical applications 
to describe rock properties.  The correction term $\beta$ can be expressed in terms of the Thomsen parameters as, 
\beq{-97}
\beta 
\approx a_1  \epsilon  + a_2 \delta + a_3\gamma, 
\eeq
where the coefficients $a_1$,   $a_2$ and  $a_3$ are 
\beq{96}
a_1 = -\frac{8a_l}{15}  - \frac{\kappa^2 a_t}{15} ,
\quad
a_2 = \frac{a_l}{15}  - \frac{\kappa^2a_t}{30} ,
\quad
a_3 = -\frac{a_t}{3}.
\eeq

\begin{figure}
				\begin{center}	
				\includegraphics[width=3.6in , height=2.6in 					]{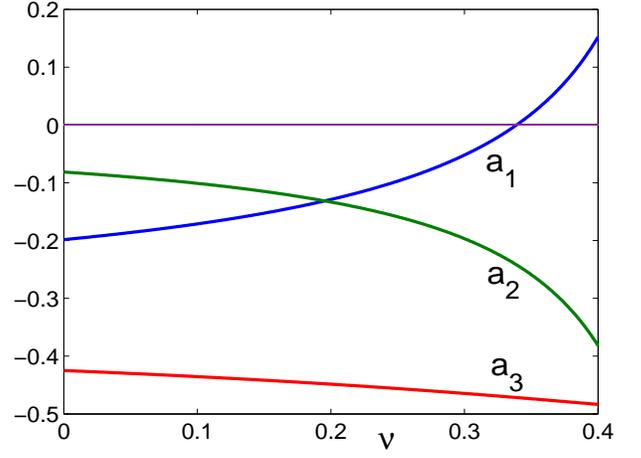} 
	\caption{The non-dimensional parameters $a_1$, $a_2$ and $a_3$ for weak transverse isotropy as a function of the Poisson's ratio $\nu$ of the background medium.}
		\label{fig2} \end{center}  
	\end{figure}

\section{Examples and discussion}\label{sec5}

The participation  matrix was computed for many   anisotropic solids.   Table II  
summarizes the results for a  selection of materials with anisotropy ranging from weak to strong.   The table provides the numerical values of diagonal elements of ${\bf G}$ (there are no off-diagonal elements for the symmetries considered).   In each case the   elements sum to three, $G_{11}+G_{22}+G_{33}=3$, although the individual numbers can differ markedly from unity.  

In order to quantify the level of anisotropy, the  table also shows the number \emph{dist}.   This is a nondimensional  positive   measure of the degree of anisotropy of a set of anisotropic elastic constants.  \emph{dist} is chosen here as the log-Euclidean distance or length  from isotropy \cite{Norris05f,moakher06b}, although other measures are possible, see Norris \cite{Norris05f} for a comparative discussion. The log-Euclidean distance has the advantage that it  is invariant regardless of whether  the compliance or stiffness tensor are  considered.  We use $dist$ as a convenient and simple measure of the degree of anisotropy.    Appendix \ref{appc} provides a little more detail on its exact definition, including a short  Matlab script to compute $dist$.

\begin{table}	\label{tab2}
\caption{The participation  matrix  ${\bf G}$ for a variety of anisotropic materials.  Sym denotes material symmetry:   transversely isotropic (TI), tetragonal (Tet) or orthotropic (Orth).  The  Frobenius (p=2) norm is used to compare 
${\bf G}$ with the isotropic result $({\bf I})$ and with the perturbation approximation $\tilde{\bf G}$ defined by  Eq. \rf{357}.   \emph{dist} is a non-dimensional and  invariant measure of the anisotropy \cite{Norris05f}, equal to zero for isotropy.  \emph{dist}$\ge 1$ signifies considerable anisotropy.  }  
\begin{ruledtabular}
\begin{tabular}{ll cccc cc} 
Material		&	Sym	&	$G_{11}$	&	$G_{22}$	&	$G_{33}$	&	$|{\bf G}\! -\! {\bf I}|$	&	$|{\bf G}\! -\! \tilde{\bf G}|$	&	\emph{dist}	 \\ 
&&&& &&&   \\  
Beryllium\footnotemark[1]		&	TI	&	1.05	&	1.05	&	0.89	&	0.13	&	0.00	&	0.22	 \\ 
Sulphur\footnotemark[1]		&	Ort	&	0.95	&	1.32	&	0.73	&	0.42	&	0.11	&	0.95	 \\ 
Cadmium\footnotemark[1]		&	TI	&	0.73	&	0.73	&	1.55	&	0.67	&	0.10	&	1.02	 \\ 
Barium titanate\footnotemark[2]		&	Tet	&	0.81	&	0.81	&	1.39	&	0.48	&	0.01	&	1.11	 \\ 
Rochelle salt\footnotemark[1]		&	Ort	&	1.38	&	0.65	&	0.97	&	0.52	&	0.09	&	1.16	 \\ 
Zinc\footnotemark[1]		&	TI	&	0.71	&	0.71	&	1.58	&	0.71	&	0.14	&	1.17	 \\ 
Graphite/Epoxy\footnotemark[3]		&	TI	&	1.38	&	1.38	&	0.25	&	0.92	&	0.81	&	2.35	 \\ 
Tellurium dioxide\footnotemark[4]	&	Tet	&	1.30	&	1.30	&	0.40	&	0.74	&	0.72	&	2.87	 \\ 
Mercurous iodide\footnotemark[4]		&	Tet	&	1.37	&	1.37	&	0.26	&	0.91	&	0.14	&	3.02	 \\ 
Spruce\footnotemark[1]		&	Ort	&	1.35	&	1.63	&	0.02	&	1.22	&	1.30	&	5.59	 \\ 
\end{tabular}
\end{ruledtabular}
\footnotetext[1]{Elastic moduli from Ref.~\onlinecite{Musgrave}.}
\footnotetext[2]{From Ref.~\onlinecite{Truell}.}
\footnotetext[3]{From Ref.~\onlinecite{Kriz96}.}
\footnotetext[4]{From Ref.~\onlinecite{Cazzani05}.}
\end{table}
  
Large deviations from the isotropic participation tensor are apparent.  Consider the ratio $R$ of the largest to smallest element of $\bf G$.   Even for small to moderate anisotropy, such as Cadmium we see that $R = G_{33}/G_{11} >2$.  The ratio becomes much larger for the more anisotropic materials considered.  Spruce is included because of its enormous ratio, $R \ge \approx 80$.  
These ratios can be compared with the results  for the relative partition of the diffuse wave energy at the free surface of an isotropic solid.  If ${\bf e}_3$ is the normal to the surface, then the calculations of Weaver \cite{Weaver85} indicate that $1\le G_{33}/G_{11} < 1.25$ where the lower (upper) bound is reached as $\nu$ approaches $1/2$ $(0)$.  The upper bound $\approx 1.25$ is approximate and based on Fig. 3 of Ref.~\onlinecite{Weaver85}.   
  
The numbers in Table II 
indicate that the  perturbation approximation is adequate for small anisotropy.  This can be characterized loosely as $0<$ \emph{dist}$\le 1$, and strong anisotropy is 
\emph{dist}$\ge 2$, roughly.   The examples in the Table suggest that the weak anisotropy approximation is not useful in the presence of strong anisotropy.  This is evident from the fact that the errors $|{\bf G}\! -\! {\bf I}|$ and 	$|{\bf G}\! -\! \tilde{\bf G}|$ are of the same order of magnitude for the strongly anisotropic materials, whereas $|{\bf G}\! -\! \tilde{\bf G}|$ is much less than $|{\bf G}\! -\! {\bf I}|$ for weak anisotropy. 

\rev{
We note that for all materials considered the numerical calculations show Eq. \rf{792}   underestimating $\tr \langle {\bd Q}^{-3/2}\rangle $.  However, the more refined perturbation expansion of  $\tr \langle {\bd Q}^{-3/2}\rangle $ by Fedorov \cite{fed} suggests that this is  not a universal result. }

The dependence of $\bd G$ and $d(\omega)$ on the  moduli is obviously complicated by virtue of the averages required in Eq. \rf{635}.  However, the formula \rf{357} for $\bd G$  for weak anisotropy illustrates the dependence more explicitly.   The form of the matrices ${\bd C}_l$ and ${\bd C}_t$ imply that only 12 combinations of the 21 independent anisotropic moduli enter into the first term in the perturbation expansion.  For orthotropic materials, with 9 independent moduli, this number reduces to 6, and the matrices ${\bd C}_l$ and ${\bd C}_t$  are then diagonal.    In the case of weak TI only two combinations of moduli influence $\bd G$, see Eq. \rf{-98}.  

The non-dimensional tensor $\bf G$ also has important implications for radiation from a point source.  The connection follows from the relation \rf{564}  between $\bf G$ and $\bf A$, combined with the correspondence between the drive point admittance tensor and the radiation efficiency in Eq. \rf{904}.   Thus, the direction in which a force must be applied to most efficiently radiate power is the  principal direction of $\bf G$ with the largest element.  Conversely, the least amount of power is radiated if the force is directed along the principal direction with the smallest element.  For instance, Table II 
indicates that a  point force of given magnitude will radiate most power in Cadmium if the force is directed along the axis of hexagonal symmetry.  The situation is reversed for aligned graphite/epoxy, where  forcing along  the fiber direction produces the least amount of total radiated power.  

The inverse problem of determining anisotropy from measurements of $\bd G$ is clearly ill-posed.  However, possible measurement could be advantageous in particular circumstances.  Consider for instance, 3-component measurement of the displacement downhole in a borehole environment.  Assuming the frequency is such that the wavelengths are large compared with the bore radius, the 3-component data is sufficient to compute the auto-correlation and hence $\bd G$.  The  principal directions of $\bf G$ and the relative magnitude of its  diagonal elements provides significant information about the local geostratigraphy and formation properties.

\section{Conclusion}
We have derived general formulas for diffuse waves in anisotropic solids.  The main results are   
concise expressions for the modal  density  per unit volume and frequency, $d(\omega)$ of 
Eq. \rf{635a}, and the participation tensor $\bf G$ of Eq. \rf{635b}.  The latter is a material constant with one or two independent constants, and with principal axes dictated by the material symmetry.     In the absence of symmetry the participation tensor defines principal axes  for diffuse wave energy distribution, and for radiation efficiency.   Calculation of 
$d(\omega)$ and $\bf G$ requires, in general, averaging over the surface of the unit sphere.  Single integrals suffice for transverse isotropy, with the important quantities given in Eq. \rf{708}.  In the case of weak anisotropy, a perturbation scheme produces explicit formulas, Eqs. \rf{792} and \rf{357}.   The main quantity in all cases is the second order averaged tensor 
$\langle  {\bd Q}^{-3/2} \rangle$.    We have illustrated the results through calculations for several materials. These display the main effects that would occur in all anisotropic solids.  In particular, the deviation $\bf G$ from the unit identity tensor can be significant.  Ratios of 2 or more for the relative magnitude of diffuse wave energy in different directions in crystals can occur under moderate levels of anisotropy, with far larger ratios possible in realistic materials.  

\section*{Acknowledgment}
I would like to thank the anonymous reviewer who pointed out  relevant work by Fedorov. 
  
\appendix

\section{Derivation of Eq. \rf{564}}\label{appa} 

\rev{
We use an argument based on a modal representation \cite{Weaver85} for the solution to the point force problem, 
\beq{901}
\big( \rho \frac{\partial^2 ~}{\partial t^2} - L\big) {\bd u} =   {\bd F} \delta( {\bd x} - {\bd x}_0) \cos \omega t, 
\eeq 
where $L$ is a second order differential operator. 
The   resulting velocity ${\bd v} = \partial {\bd u}/ \partial  t$ may  be found by standard means as  
\beq{903}
{\bd v} = \frac1{\rho}\Re \sum\limits_m \frac{-i\omega {\bd F}\cdot {\bd u}_m({\bd x}_0)\,  {\bd u}_m({\bd x})
}{\omega_m^2 - \omega^2 - i0} \, e^{-i\omega t}, 
\nonumber 
\eeq
where the modes ${\bd u}_m({\bd x})e^{-i\omega_m t}$ are   solutions of the homogeneous equation \rf{901}, with the properties
\begin{align}
& \delta( {\bd x} - {\bd x}_0) {\bd I} = \sum\limits_m {\bd u}_m({\bd x}){\bd u}_m({\bd x}_0),
\nonumber \\ 
& \int_V \dd {\bd x}\,  {\bd u}_m({\bd x})\cdot {\bd u}_m({\bd x}) = 1. \nonumber
\end{align}
The  power output averaged over a cycle is therefore
\bal{-55}
\Pi ({\bd x}_0, \omega ) =& \frac{\omega}{2\pi}\int_0^{{2\pi}/{\omega}} \dd t \cos \omega t \, {\bd F}\cdot{\bd v} ({\bd x}_0,t)
\nonumber \\ 
=  & \frac{ 1}{2\rho} \sum\limits_m    [{\bd F}\cdot {\bd u}_m({\bd x}_0)]^2
\Re \frac{-i\omega}{\omega_m^2 - \omega^2 - i0} .
\end{align}
}

\rev{
 The strict non-dissipative limit of 
$\Re [{-i\omega}({\omega_m^2 - \omega^2 - i0})^{-1}]$ is  
$\pi \omega  \delta( \omega_m^2 - \omega^2 ) = \frac12 \pi  \delta( \omega_m - \omega )$ where $\delta$ is the Dirac delta function.  However, modal overlap in the presence of non-zero dissipation spreads the influence over many modes.  The effect is to make 
$\Re [{-i\omega}({\omega_m^2 - \omega^2 - i0})^{-1}] \rightarrow 
\frac12 \pi  f( \omega_m - \omega )$ where $f(\nu )$ is smooth with bounded support in $\nu \in \{ -\Omega, \Omega\}$, say,  and unit sum: 
\beq{-44}
\sum_{\omega_m} '  f( \omega_m - \omega  ) =1.
\eeq 
Here $\sum_{\omega_m}'$ indicates the sum over modal frequencies 
$\omega_m\in \{ \omega-\Omega ,     \omega+ \Omega \}$. 
Using  the density of modes, $V d(\omega_m)$, to replace the sum over modes in \rf{-55} by a sum over modal frequencies,  gives
\beq{+11}
\Pi ({\bd x}_0, \omega ) = \frac{\pi V}{4\rho} 
\sum_{\omega_m}'
d(\omega_m  )
f( \omega_m - \omega  ) [{\bd F}\cdot {\bd u}_m({\bd x}_0)]^2.
\eeq
We now make the  assumption that the support of $f(\nu )$ is small enough  that 
the modal density function, $d(\omega_m  )$, may be replaced by $d(\omega  )$.  This is perfectly reasonable based on known forms for $d(\omega  )$, 
 e.g. Eq. \rf{86}.  
At the same time, we assume that the support of $f(\nu )$ is sufficiently large that
we may use the equipartition of energy among modes to make the replacement (see Eq. \rf{560}) 
\beq{+12}
V \sum_{\omega_m}'
f( \omega_m - \omega  ) {\bd u}_m  \otimes {\bd u}_m  
\rightarrow V \frac{\rho \omega^2}{E} \bar{\bd u}   \otimes \bar{\bd u}  
= \frac13 {\bd G}.
\eeq
Hence, 
\beq{922}
\Pi ({\bd x}_0, \omega ) 
= \frac{\pi }{12\rho}  d(\omega) \, {\bd F}\cdot {\bd G}\cdot {\bd F}, 
\eeq
and since $\bd F$ is arbitrary, the admittance $\bd A$  follows from  the definition of $\Pi $ in \rf{904}.  
This completes the derivation of the identity \rf{564}.
}

\section{Derivation of Eqs.  \rf{-21} and \rf{29}}\label{appb}

The Cayley-Hamilton relation for ${\bd Q}$ is $p({\bd Q}) = 0$, where $p$ is the characteristic cubic polynomial defined in  Eq. \rf{+33}, 
and  $I_1( {\bd n})$, $I_2( {\bd n})$, $I_3( {\bd n})$ are the invariants define in Eq. \rf{4}.  
Thus, 
\beq{5}
 I_1 = \lambda_1 +\lambda_2 +\lambda_3,
\, 
   I_2 = \lambda_1 \lambda_2+\lambda_2 \lambda_3+\lambda_3\lambda_1,
  \, 
  I_3 = \lambda_1 \lambda_2 \lambda_3,  \nonumber 
\eeq
and since $\lambda_\alpha =   v_\alpha^2$, it follows that the invariants are all positive, 
$I_1 >0$, $I_2 >0$ and $I_3 >0$.
Multiplying \rf{-33} by ${\bd Q}^{-1}$ and ${\bd Q}^{-2}$ yields  equations for the same quantities: 
\begin{subequations}
\begin{align}\label{6}
{\bd Q}^{-1}  &= I_3^{-1}{\bd Q}^2 - I_1 I_3^{-1}{\bd Q} +I_2 I_3^{-1} {\bd I},
\\
 {\bd Q}^{-2}  &= I_3^{-1}{\bd Q} - I_1 I_3^{-1}{\bd I} +I_2 I_3^{-1} {\bd Q}^{-1}.
\end{align}
\end{subequations}
Eliminating ${\bd Q}^{-1}$ gives an equation for $ {\bd Q}^{-2} $:
\beq{7}
{\bd Q}^{-2}  = I_3^{-2}\big[ I_2 {\bd Q}^2 - ( I_1I_2-I_3) {\bd Q} + ( I_2^2-I_1I_3){\bd I} \big].
\nonumber 
\eeq
We next derive a similar type of  equation for $ {\bd Q}^{1/2}$ using a method due to Hoger and Carlson \cite{Hoger84}. The product of this with ${\bd Q}^{-2}$, combined with the Cayley-Hamilton equation \rf{-33} yields the desired relation  \rf{-21}. 

First we note the general expression
\bal{801}
& ( {\bd Q} - \lambda{\bd I})^{-1} = 
\nonumber \\ & \quad \frac{1}{p(\lambda)}\big[ 
- {\bd Q}^2 + (I_1- \lambda){\bd Q} - ( \lambda^2 - I_1 \lambda +I_2){\bd I}\big],
\end{align}
where $p$ is the characteristic polynomial for ${\bd Q}$, from Eq. \rf{+33}. 
The identity \rf{801} may be checked by direct multiplication and use of Eq. \rf{-33}. 
The square root tensor  ${\bd R} \equiv  {\bd Q}^{1/2}$ satisfies the Cayley-Hamilton equation 
\beq{9}
{\bd R}^3 - i_1 {\bd R}^2 +i_2 {\bd R} - i_3 {\bd I}= 0, 
\eeq
where $i_1$, $i_2$ and $i_3$ are related to the invariants of $\bd Q$ by 
\beq{12}
I_1 = i_1^2-2i_2,
\quad
I_2= i_2^2 - 2i_1 i_3, 
\quad
I_3= i_3^2. 
\eeq
Explicit formulae for $i_1 $, $i_2$ and $i_3$ are given in \rf{13}. Rearranging \rf{9} as
${\bd R} ( {\bd R}^2 + i_2 {\bd I} )  = i_1 {\bd R}^2 + i_3 {\bd I}$ 
and using ${\bd R}^2 = {\bd Q}$ gives
\beq{11}
{\bd R}   = (i_1 {\bd Q} + i_3 {\bd I})( {\bd Q} + i_2 {\bd I} )^{-1}. 
\eeq
Application of \rf{801} along with some simplifications using \rf{12}, such as
$p(-i_2) = - (i_3- i_1 i_2)^2$, yields
\beq{802}
 {\bd Q}^{1/2} =  (i_3- i_1 i_2)^{-1}\, \big[  {\bd Q}^2 + (i_2-i_1^2) {\bd Q} - i_1i_3 {\bd I}\big]. 
\eeq
Combining Eqs. \rf{7} and \rf{802} gives Eq. \rf{-21}. 
 Alternatively,
\beq{-2}
 {\bd Q}^{-3/2} =    a{\bd Q}^2 + b{\bd Q} + c {\bd I},
 \eeq
 where
 \bal{-366}
a&=  \frac{I_3 (i_2-   i_1^2) - I_2 i_1 i_3  }{ I_3^2  (i_3- i_1 i_2) },
\nonumber 
 \\ 
 b&=
 \frac{
   I_1  I_3 (i_1^2- i_2) + (I_1  I_2 - I_3) i_1 i_3
 }{ I_3^2  (i_3- i_1 i_2) },
 \\ 
 c&=  
  \frac{
 I_3^2  + I_2  I_3(i_2  -   i_1^2 )  + ( I_1  I_3  - I_2^2) i_1 i_3 
 }{ I_3^2  (i_3- i_1 i_2) }.
 \nonumber 
\end{align}

The second form \rf{29} for  ${\bd Q}^{-3/2}$ is based on the identity \rf{2}.  The tensor products of eigenvectors for $\lambda_i$ satisfy 
\beq{071}
 {\bd q}_i  \otimes {\bd q}_i = \frac{
({\bd Q}- \lambda_j {\bd I})  ({\bd Q}- \lambda_k {\bd I})   }
{ (\lambda_i - \lambda_j) (\lambda_i - \lambda_k)},\, i\ne j\ne k\ne i\quad\text{(no sum)}. 
\nonumber 
\eeq
This follows, for example, by eliminating the other two tensor products using the spectral expressions for ${\bd I}$, ${\bd Q}$ and ${\bd Q}^2$.  The dependence on 
$  \lambda_j$ and $\lambda_k$ can be removed in favor of $  \lambda_j$ and the invariants 
$I_1$ and $I_3$, and hence Eq. \rf{291}. Note that the latter can be expressed
\beq{294}
{\bd N}(\lambda , {\bd n})  = \frac{1}{\lambda p'(\lambda)}
\big[ \lambda {\bd Q}^2 + (\lambda - I_1) \lambda  {\bd Q} + I_3{\bd I} \big],
\eeq
where $p'(x)$ is the derivative of the characteristic polynomial.  This indicates that the general expression \rf{291} is invalid at double roots where the slowness surface exhibits degeneracy, and proper limits are required.   The possibility of such points  does not present  a practical impediment to numerical integration.

\section{The log-Euclidean distance}\label{appc} 

The procedure \cite{Norris05f} is to first calculate an effective isotropic set of moduli analogous to 
$c^{(0)}_{ijkl}$ of Eq. \rf{000} but for the matrix logarithm of the 6-dimensional Voigt matrix of moduli $C_{IJ}$.  Some matrix factors are required to convert from the Voigt notation.  The following Matlab lines compute \emph{dist} if $C$ is the 6$\times$6 Voigt matrix.  
\begin{verbatim}
J = 1/3*[1 1 1 0 0 0]'*[1 1 1 0 0 0] ; 
K = eye(6)-J; 
T = diag([ 1 1 1 sqrt(2)*[1 1 1] ]);
L = logm(T*C*T);
dist = norm(logm( J*exp(trace(J*L)) 
    + K*exp(1/5* trace(K*L)) )- L ,'fro');
\end{verbatim}



\end{document}